\documentclass[12pt]{JHEP3}

\pdfoutput=1
\usepackage{bm}
\usepackage{epsfig}
\usepackage{citesort}
\usepackage{graphicx}
\usepackage{multirow}

\def \be  {\begin{equation}}
\def \ee  {\end{equation}}
\def \ber  {\begin{eqnarray}}
\def \eer  {\end{eqnarray}}

\newcommand{\om}{\Omega_m}
\newcommand{\ode}{\Omega_{de}}

\title{Testing cosmological models using relative mass-redshift abundance of SZ clusters} 

\author{Arman Shafieloo\\
	Institute for the Early Universe, Ewha Womans University\\ 
	Seoul, 120-750, South Korea\\
	E-mail: \email{arman@ewha.ac.kr}}

\author{George F. Smoot\\
       Lawrence Berkeley National Laboratory, Berkeley, CA 94720, USA\\
       \emph{and}\\
       Institute for the Early Universe, Ewha Womans University\\
       Seoul, 120-750, South Korea\\
       \emph{and}\\
       Paris Centre for Cosmological Physics, Universite Paris Diderot, France \\
       E-mail: \email{gfsmoot@lbl.gov}}


\abstract{Recent detection of high-redshift, massive clusters through Sunyaev-Zel'dovich observations has opened up a new way to test cosmological models. 
It is known that detection of a single supermassive cluster at a very high redshift can rule out many cosmological models all together. 
However, since dealing with different observational biases makes it difficult to test the likeliness of the data assuming a cosmological model, most of the cluster data (except those with high mass-redshift) stays untouched in confronting cosmological models with cluster observations. 
We propose here that one can use the relative abundance of the clusters with different masses at different redshifts to test the likeliness of the data in the context of cosmological models. For this purpose we propose a simple parametric form for the efficiency of observing clusters at different mass-redshift and we test if the standard $\Lambda$CDM model can explain the observed abundance of the clusters using this efficiency parameterization. We argue that one cannot expect an unusual and highly parametric form of the efficiency function to fit the observed data assuming a theoretical model. Using many realizations of Monte Carlo simulations we show that the standard spatially flat $\Lambda$CDM model is barely consistent with the SPT cluster data using a simple and plausible two-dimensional efficiency function for detection of the clusters. More cluster data are needed to make any strong conclusion.}  

\begin{document}



\section{Introduction}

A key task of modern cosmology is to determine the actual model of the universe and its parameters. There have been many efforts in last two decades using various observations to distinguish between cosmological models and put constraints on some of the key cosmological quantities like matter density, curvature and dark energy. Cosmic microwave background (CMB), supernovae type Ia, large scale structure of the universe and distribution of the galaxies and weak/strong gravitational lensing have been so far the main sources of information for cosmologists to probe and study the universe. 
However, with higher quality data, advancements in computational analysis, and more sophisticated statistical techniques, we can look for more observational sources from our surrounding universe to study the cosmos.    

Recent developments in CMB observations provides us with high angular resolution, large area mapping, 
 which can be used to detect massive clusters up to very high redshifts through  the Sunyaev-Zel'dovich (SZ) effect~\cite{SZ_review}.  
 
Clusters can be a strong discriminator against or among cosmological models. 
It takes time for the universe to become a host for massive clusters of galaxies so the time (redshift) that clusters appear in the universe is important. The predicted distribution of the mass of clusters with redshift  is related to the assumed theoretical model and also the initial conditions for the primordial fluctuations. 
Hence observations of massive clusters at different redshifts can be used to test different cosmological models. 
For instance, detection of a single supermassive cluster above a certain redshift can simply rule out those cosmological models that do not and cannot predict an abundance of such massive clusters at those redshifts. 
As an example, if we observe only one cluster with the mass of around $3 \times 10^{15}M_{Sol}$ at redshift of $z \buildrel{>}\over{\sim} 1$ the standard model of cosmology, spatially flat $\Lambda$CDM model with power law form of the primordial spectrum of the fluctuations would be ruled out. 
In other words, it is impossible for the standard model to generate such a massive cluster at such early stage of the universe (look at Figure~5 of~\cite{SPT}). 

Abundance of the massive clusters is related to the initial condition of the fluctuations too. 
Large amplitude of the initial fluctuations or large primordial non-Gaussianities could allow the universe to host few super massive clusters within our observational horizon at very high redshifts in contrast to the standard model that is unable to do so (assuming a power law form of the primordial spectrum arising from a single field slow role inflationary scenario). 
So observation of the massive-high redshift clusters can easily discriminate among different models of cosmology. 
Not only can high mass, large redshift clusters rule out class of models but also can constrain parameters of the assumed models. 
For the standard model of cosmology having lower or higher matter density can result in slower or faster growth of fluctuations in the large scale structure of the universe and this in turn affects the abundance of the massive clusters versus redshift. 

As another example clusters can be used to constrain early dark energy models where non-clustering dark energy has a considerable share of the total energy budget of the universe even at high redshifts. 
Matter density and expansion rate of the universe for these early dark energy models are different from the models with negligible dark energy at high redshifts. 
These differences can affect the growth of fluctuations and hence abundance of the massive clusters. 
So one can see that clusters can play an important role in precision cosmology and soon with more number of clusters at high redshifts (from Planck or other SZ surveys) this role will be more substantial.

Using the current cluster data (with less than 50 clusters overall from all surveys) there have been some first attempts to falsify cosmological models through exclusion curves which put an upper limit on the mass of the clusters expected in a particular model at different redshifts. 
Usually in these works only clusters with the highest mass at any specific redshift are used to falsify the cosmological models and the other clusters with lower mass have almost no role in these tests. 

In this paper we generalize usage of the cluster data proposing a method to test the likeliness of the whole data set assuming a cosmological model along with a two-dimensional efficiency function. 
In other words we check whether the observed data and the relative abundance of the clusters as a function of cluster mass and redshift can be realized with reasonable likelihood for a model of cosmology, 
if we assume a flexible efficiency function for the observation of the clusters. 
We discuss how we choose the form of the two dimensional efficiency function and we define a likelihood term for the whole data using characteristics of the Poisson distribution. Regards the background cosmological model, in this paper we have limited ourselves to the standard model, spatially flat $\Lambda$CDM model with power law form of the primordial spectrum but in principle any other model of cosmology or any other form of the primordial fluctuation can be used.  

\section{Data and Theoretical Expectations} 

In confrontation of the theoretical models with observations it is good to derive quantities for the theoretical model very close to the actual form of the observed data. 
In this way one avoids introducing additional biases in the process of error-propagation, 
while changing the form of the data. In this case of study we are dealing with clusters and the data is in the form of mass-redshift of these objects. 
Of course derivation of the mass and redshift of the clusters involves much of analytical work, 
but here we assume that we can trust the data given to us by the observers and consider it to be cosmologically model independent. 

So following this strategy we need to derive the theoretical expectation of abundance of the clusters in mass-redshift space given the cosmological model along with the initial condition of the fluctuations. 
Abundance of the clusters can be derived having the halo mass function and the comoving volume of the universe at different redshifts. 
Halo mass function relates the change in the number of clusters to the change in their mass and is a function of the present matter density, linear growth function and $\sigma_8$~\cite{tinker}. 
Comoving volume can be derived having the expansion rate of the universe. 
So setting clearly the background model one can derive the mass-redshift abundance of the clusters. 
Here we use a fitting formula given in~\cite{mortonson} with some corrections/modifications\footnote{The equation A3 in~\cite{mortonson} does not have a monotonic behavior at low redshifts. Since this equation is to provide the cumulative number of clusters above some certain redshift-mass, it should not have a maximum value at $z>0$. To correct this we assume that at low redshifts, where Hubble law holds, the expected number of clusters is directly related to the cosmic volume.} to estimate the median number of the clusters expected above a given mass and redshift for spatially flat $\Lambda$CDM model:

\begin{figure*}[!t]
\centering
\begin{center}
\vspace{-0.05in}
\centerline{\mbox{\hspace{0.in} \hspace{2.1in}  \hspace{2.1in} }}
$\begin{array}{@{\hspace{-0.3in}}c@{\hspace{0.3in}}c@{\hspace{0.3in}}c}
\multicolumn{1}{l}{\mbox{}} &
\multicolumn{1}{l}{\mbox{}} \\ [-4.2cm]
\hspace{-0.3in}
\includegraphics[scale=0.42, angle=0]{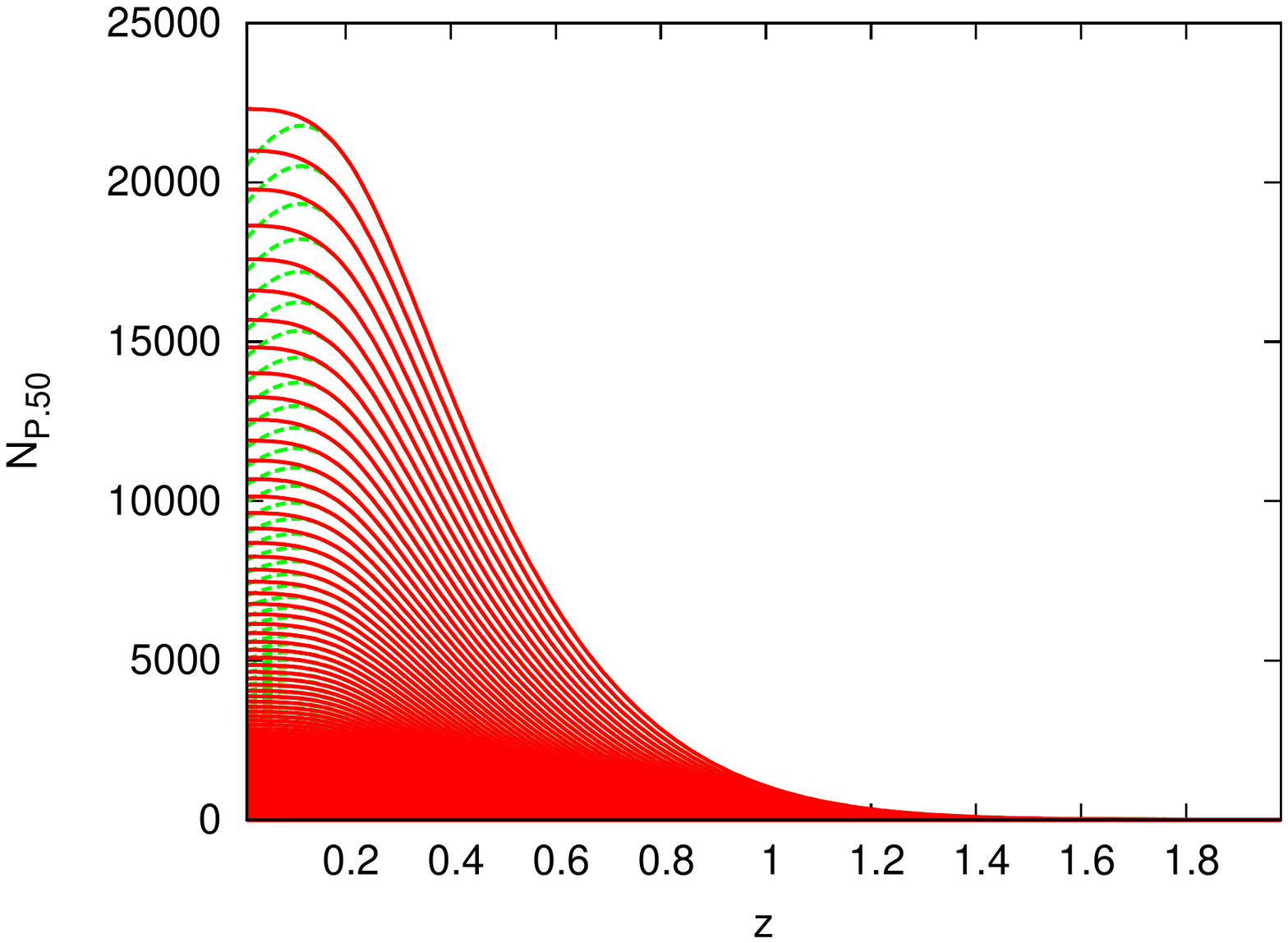}
\hspace{-1.5in}
\includegraphics[scale=0.42, angle=0]{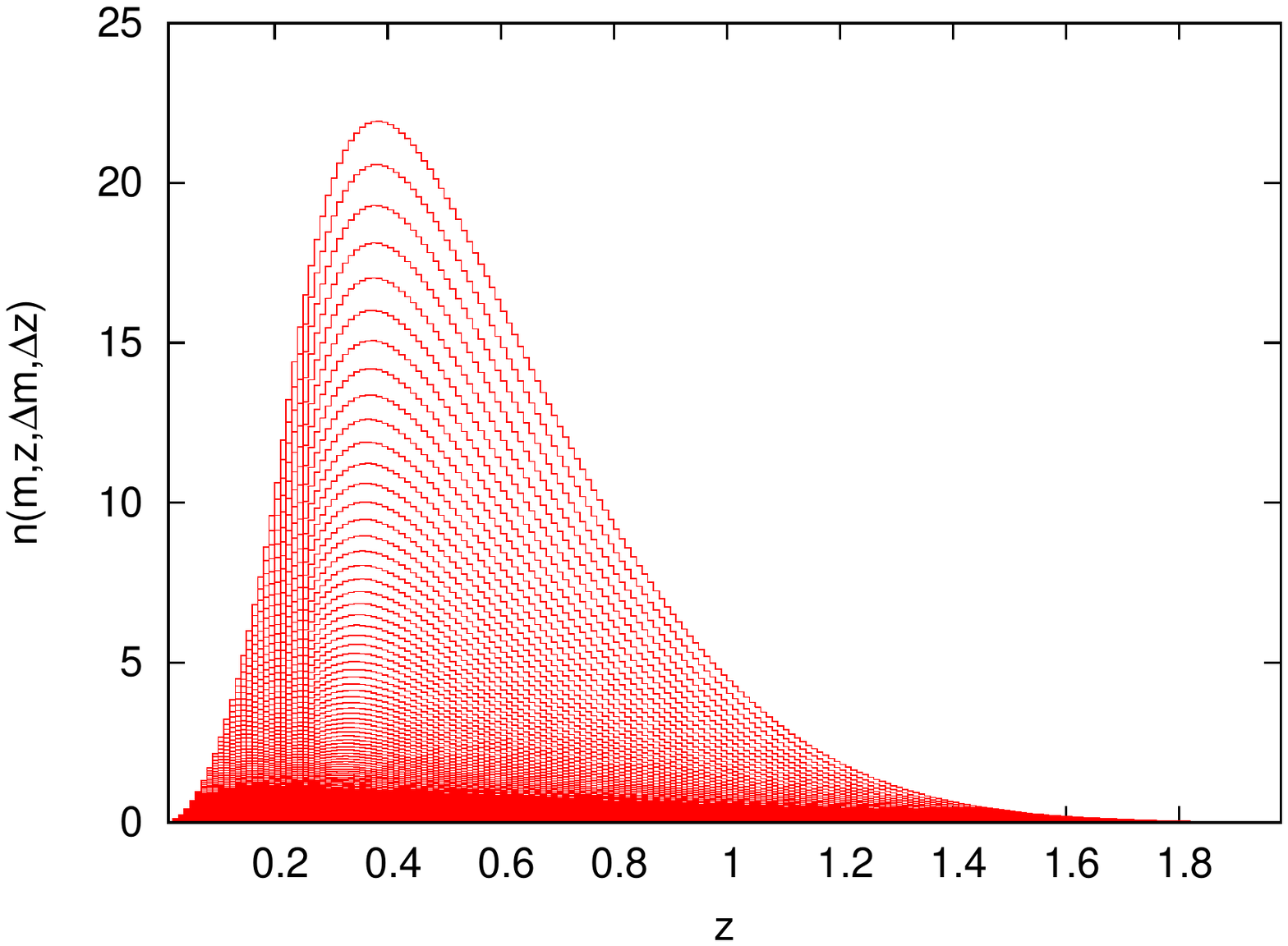}
\hspace{-0.8in}
\end{array}$
\vspace{-0.7in}
\end{center}
\caption {\small 
In the left panel we see the median number of clusters we expect to observe above redshift 
z for different masses in the whole sky using the modified fitting formulas Eq. 2.1 and 
Eq. 2.2 (red lines). Green lines are from  the fitting formula Eq. 2.1 at all redshifts 
where there is a problem at low redshifts and the fitting formula has a maximum at $z>0$. In the right 
panel we see the abundance of clusters in each redshift interval for different masses. We used 
$\Delta z=0.01$ and $\Delta m=10^{13}M_{Sol}$ as the resolutions in redshift and mass. Highest lines in 
both figures represent $m=5 \times 10^{14}M_{Sol}$. Lower lines are higher masses.}
\label{fig:fit_formula}
\end{figure*}


\ber  
Log \bar N(m,z) = 7.65 [1-exp[\alpha(z)(m-\beta(z)]] \nonumber \\ 
\alpha(z)=1.06 - 0.17 exp(-1.3z) \hspace{15 mm} \nonumber \\
\beta(z)=15.565 -0.1 Log [7.1+10^{(5.25 z)}] \nonumber \\
z \ge 0.2 \hspace{30 mm}
\label{eq:abhz}
\eer
and
\ber
\bar N(m,z) = \bar N(m,z=0.2) + [\frac{\delta N}{\delta z}]_{z=0.2} \int_z^{z=0.2} \frac{s^2}{0.2^2} ds \nonumber \\
z < 0.2 \hspace{30 mm}
\label{eq:ablz}
\eer
where $m=Log[M/(h^{-1}M_{Sun})]$. 
This simple fitting formula gives the cumulative number of clusters observed versus mass and redshift, which are the critical parameters to confront a theoretical model to the data. 
Using this fitting formula and having the boundary conditions that at very high redshifts and very large masses the derivatives of this function with respect to mass and redshift are zero, 
we can reconstruct the number density abundance of the clusters at any redshift-mass interval, $n(m,z,\Delta m,\Delta z)$, by taking two dimensional numerical derivatives in both mass and redshift directions. $n(m,z,\Delta m,\Delta z)$ gives the number of clusters in the $m \pm \frac{\Delta m}{2}$ and  $z \pm \frac{\Delta z}{2}$ intervals. 
We set $\Delta z=0.01$ and $\Delta m=10^{14}M_{Sol}$ (to test the method we set $\Delta m=10^{13}M_{Sol}$) as the resolutions in our numerical calculations; however, probability results are independent of these values. We should also note that though setting $z=0.2$ seems to be an arbitrary choice as a boundary between Eq.~\ref{eq:abhz} and Eq.~\ref{eq:ablz}, results would not change if we set this boundary redshift to any other value around $0.2$. One could also work out a new modified fitting formula similar to Eq.~\ref{eq:abhz} but it is not our main concern in this paper.

In Figure.~\ref{fig:fit_formula} left panel we see the median number of clusters we expect to observe above some certain redshift for different masses in the whole sky using the above modified fitting formulas Eq.~\ref{eq:abhz} and Eq.~\ref{eq:ablz} (red lines). Green lines are from  the fitting formula provided by~\cite{mortonson} (using Eq.~\ref{eq:abhz} at all redshifts). In the right panel of Figure.~\ref{fig:fit_formula} we see the abundance of clusters in each redshift interval for different masses. We used $\Delta z=0.01$ and $\Delta m=10^{13}M_{Sol}$ as the resolutions in redshift and mass. Highest lines in both figures represent $m=5 \times 10^{14}M_{Sol}$ and lower lines are for higher masses.



\begin{figure}[!h]
\includegraphics[width=0.7\columnwidth]{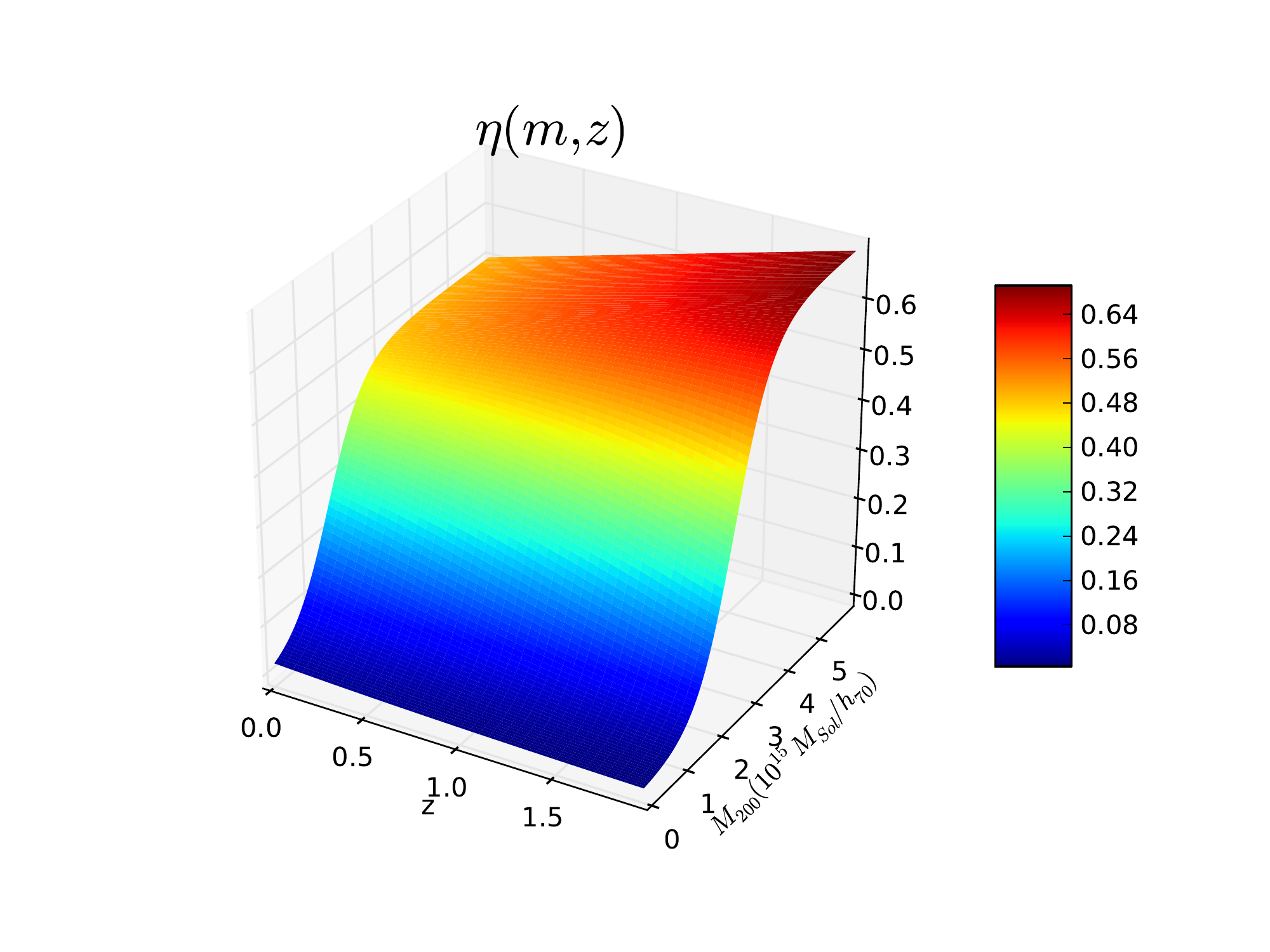}
\caption {\small A schematic form of the efficiency function $\eta(m,z)$ from Eq. 3.9 for some arbitrary choices of the free parameters.}
\label{fig:schem}
\end{figure}


\begin{figure*}[!t]
\centering
\begin{center}
\vspace{-0.05in}
\centerline{\mbox{\hspace{0.in} \hspace{2.1in}  \hspace{2.1in} }}
$\begin{array}{@{\hspace{-0.3in}}c@{\hspace{0.3in}}c@{\hspace{0.3in}}c}
\multicolumn{1}{l}{\mbox{}} &
\multicolumn{1}{l}{\mbox{}} \\ [-2.8cm]
\hspace{-0.4in}
\includegraphics[scale=0.4, angle=0]{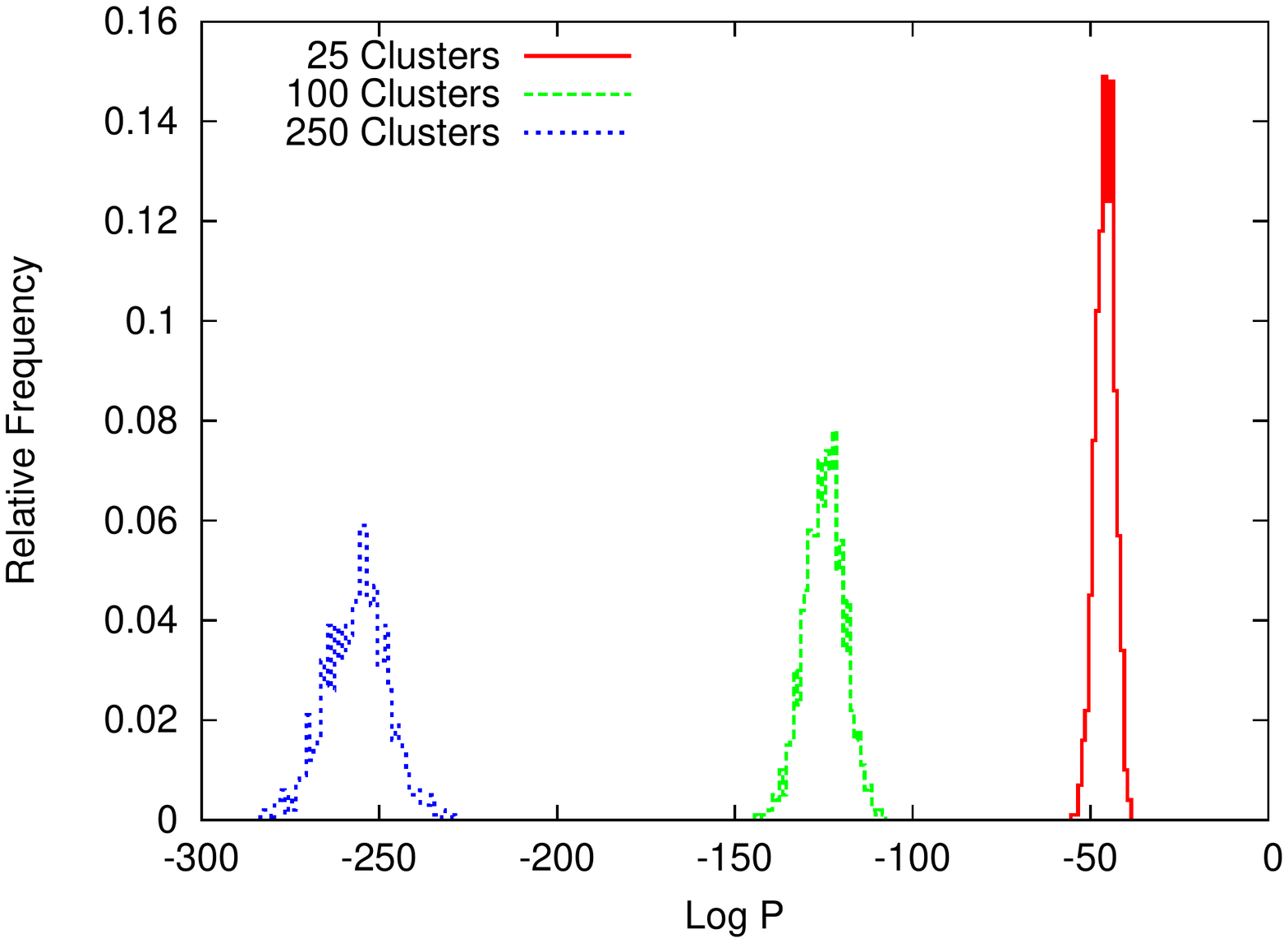}
\hspace{-1.2in}
\includegraphics[scale=0.4, angle=0]{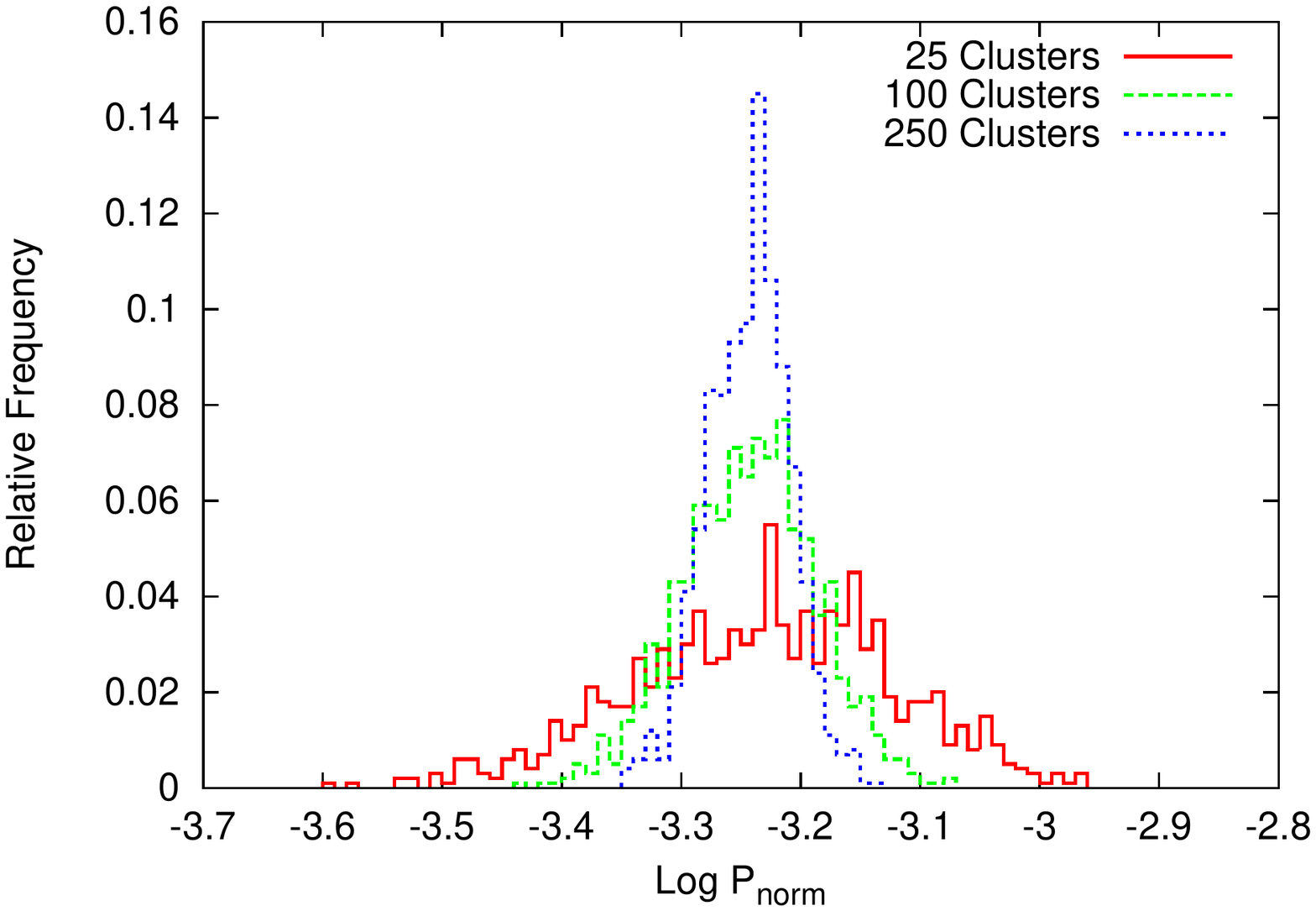}
\hspace{-1.1in}
\vspace{-0.2in}
\end{array}$
\vspace{-0.2in}
$\begin{array}{@{\hspace{-0.3in}}c@{\hspace{0.3in}}c@{\hspace{0.3in}}c}
\multicolumn{1}{l}{\mbox{}} &
\multicolumn{1}{l}{\mbox{}} \\ [-2.8cm]
\hspace{-0.4in}
\includegraphics[scale=0.4, angle=0]{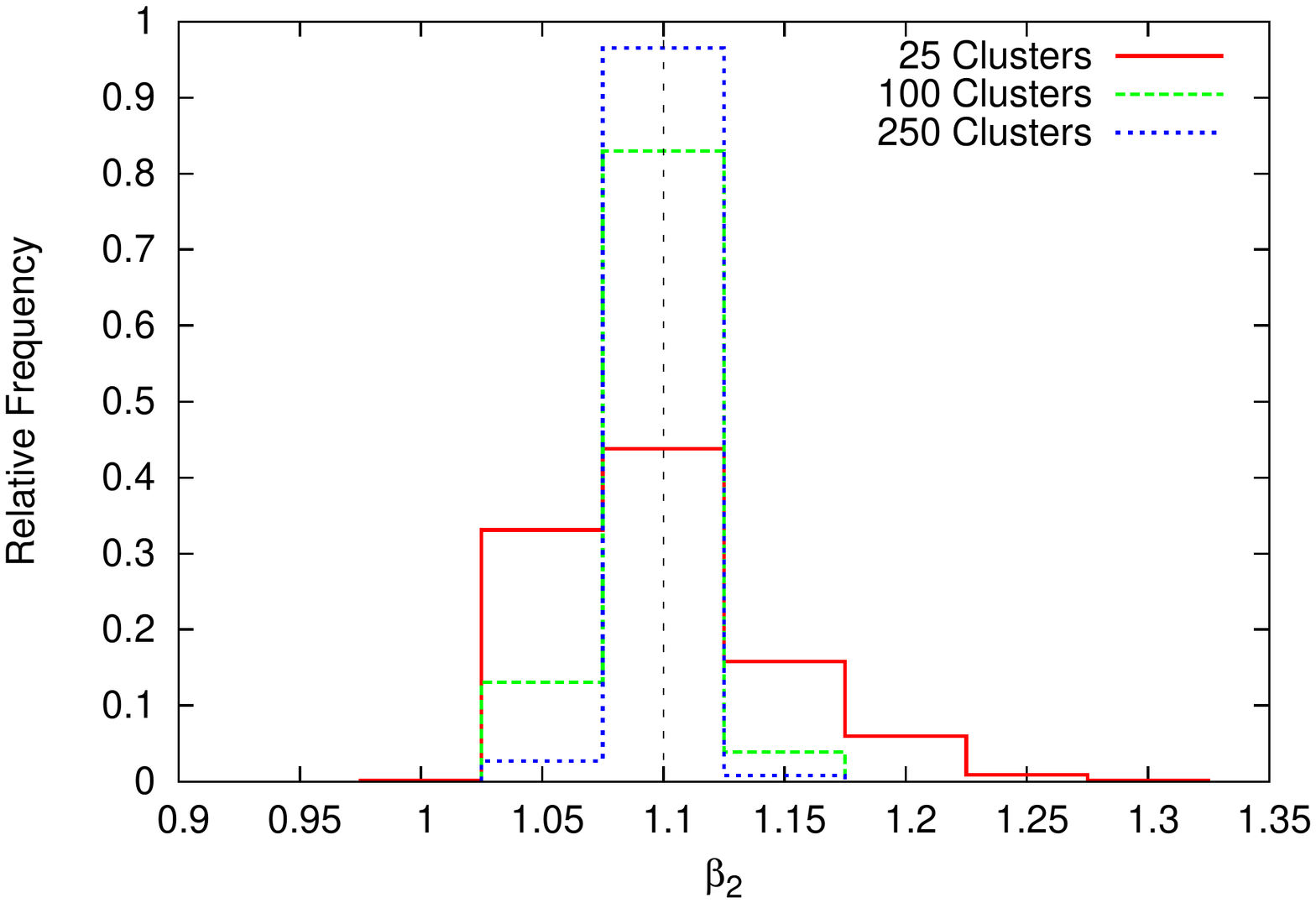}
\hspace{-1.2in}
\includegraphics[scale=0.4, angle=0]{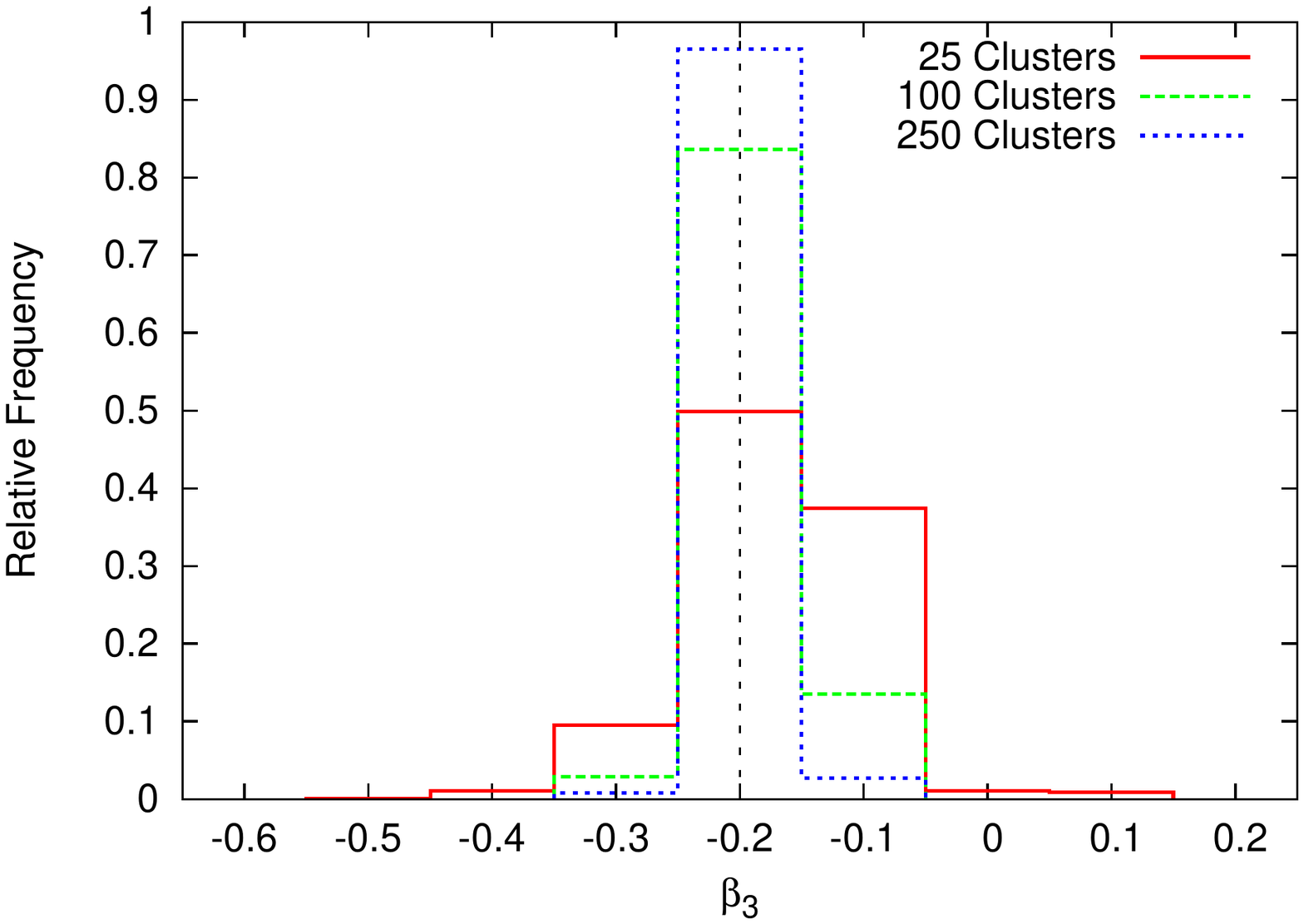}
\hspace{-1.1in}
\end{array}$
\vspace{-0.3in}
\end{center}
\caption {\small Top panels: Relative frequency of the likelihood function Eq. 4.6 (left) and normalized likelihood function Eq. 4.7 (right) for different data sizes. One can see that for more number of data points normalized likelihood confidence limits shrinks closer to a mean value. These results are generated using 1000 Monte Carlo data realizations. Lower panels: Relative frequency of the best fit derived values of the efficiency parameters for different sizes of the data. Dotted vertical lines represents the assumed fiducial values of the efficiency parameters which we simulated the data based on them. One can see that the fiducial model can be recovered quite accurately using our proposed method.}  
\label{fig:Sim}
\end{figure*}


\section{Detection Efficiency} 

We must relate the theoretical expectations to the observed data by defining an efficiency function $\eta(m,z)$. 
The efficiency function $\eta(m,z)$ estimates what fraction of the clusters can be observed due to observational-technical limitations. One does not need a precise detection/observation efficiency but only needs general principles and a limited number of plausible free parameters that can be included in the fitting. Then one can expect to get reasonably accurate answers for fitting, compare to estimates of efficiency and see that they are plausible. 

One is used to distant objects being more difficult to detect; however,
the SZ effect is an effective cosmological probe because its brightness change is independent of redshift.
The SZ signal along the line of sight is set by the Comptonization factor 
\be
y = \int n_e(r) \sigma_T \frac{k_B T_e(r)}{m_e c^2} d l =  \frac{\sigma_T}{m_e c^2} \int p_e d l
\ee
where $p_e$ is the pressure of the electrons which is proportional to the mass of the cluster.

However, the total SZ signal depends upon the cluster's apparent solid angle on the sky.
That is we must integrate over all the lines of sight through the cluster.
The apparent solid angle of a given cluster decreases with the inverse square of its angular diameter distance 
until about a redshift $z \sim 1$ where it begins to rise again because of the additional $1 + z$ factor in favor of the angular diameter distance. \footnote{The angular diameter distance, $D_A$, the proper motion distance, $D_{PM}$ and the luminosity distance, $D_L$ are simply related in any metrical geometry
\be
D_A = \frac{D_{PM} }{ (1 +z)} = \frac{D_L}{ (1 + z)^2} 
\ee
where in general and in presence of curvature and dark energy we have
\be
D_{PM}(z)=\frac{1}{\sqrt{1-\om-\ode}}\sinh\left(\sqrt{1-\om-\ode}\int_0^z \frac{dz'}{h(z')}\right)\,. \label{eq:dl}
\ee
and
\be
h(z)^2\equiv[H(z)/H_0]^2=\om (1+z)^3 + (1-\om-\ode)(1+z)^2+\ode \exp\left[3\int^z_0 \frac{dz'}{1+z'}\, [1+w(z')]\right]
\ee}

The effective cluster solid angle, $\Omega_{CSA}$, is approximately its FWHM subtended angle squared , 
or
\ber
\Omega_{CSA}  &\sim& (\pi /4) \theta_{FWHM}^2 \cr
& \propto& \left( size / D_A \right)^2 \cr  
 & \propto & \left( Mass^{1/3}  (1 + z) /  S_k \right)^2
\eer
where  in the last expression the linear size of the cluster is assumed to go with the cube root of the mass, whereas in reality it is a bit more complicated. 

\ber
S_k(z) &=& \int_0^z \frac{c}{H(z')} d z' \cr
&=& \frac{c}{H_0} \int_0^z \frac{d z}{\sqrt{\Omega_m (1 + z)^3 +(1-\Omega_m)}}
\eer

where the second expression is for the standard model of cosmology, the flat $\Lambda$CDM model of the universe.

If we ignore the complications about mass versus linear scale size of clusters, 
and remember the additional factor of mass in the SZ signal, we
then would find for a given signal threshold, presumably set by signal to noise,
we would have a cluster mass threshold that depends upon redshift roughly as
\footnote{An estimate~\cite{Chamballu2010} of the SZ signal is 
\be
= 2.87 \times 10^{-3} arcmin^{2} \left[ \frac{M_{500}}{ 3 \times 10^{14} h_{70}^{-1} M_{Sol}} \right]^{5/3} E(z)^{2/3} \left[ \frac{500 Mpc}{D_A(z)} \right]^2
\ee
where $E(z \equiv H(z)/H_0$, $M_{500}$ is the cluster mass within 500 MPc, and $D_A(z)$ is the angular diameter distance to redshift $z$.}
\be
m_{threshold} \propto  \left( S_k / (1+z) \right)^{3/5}
\ee
which has the property for the $\Lambda$CDM  current model of peaking at a redshift of about $z \sim 1.5$
and decreasing on either side with a roughly asymmetric parabolic shape.

A second effect is that for clusters at higher redshift, the confusion noise is less, except for the otherwise constant CMB primordial fluctuations. 
One of the largest reasons for this is though the SZ effect has surface brightness that is independent of redshift,
the confusing signal from the sources in the cluster falls of with the luminosity distance squared. 
Here the factor of $(1 + z)$ difference between angular diameter distance and luminosity distance
means that these confusion signals drop off with an extra factor of $(1 + z)^2$. 

In addition to these two factors there may be some additional observational or instrumental effects that confuse things 
though the SZ effect is special in being so independent of redshift.
We chose a reasonable function for the redshift dimension of the efficiency parameterization based in part on these expectations and that we can test the effectiveness of this through the fitting process.
We know that the observational efficiency for detection of the low mass clusters must be low simply because their signal is not sufficient to be detected. 
So our efficiency function in the mass dimension must allow a high efficiency at high masses and low efficiency at low masses providing a smooth transition between the two. 
At the same time there must be a saturation for the efficiency function at the two ends as it cannot be lower than zero and higher than $1$. 
Within algebraic functions $\tanh$ has such behavior that smoothly connects a minimum and a maximum value monotonically and in the extreme cases can act as a step function. 
So by this justifications we use $\tanh$ to define our efficiency function. 

Our efficiency fitting function is
\be
\eta(m,z) = A \frac{\tanh [\beta_{1}(m-\beta_{2}+\beta_{3}z + \beta_{4} z^2)]+1}{2} 
\label{eq:effparam}
\ee

where the $\beta_3$ and $\beta_4$ should help handle the mass function outlined above.\footnote{There might be even a better form that give the rough mass function shape with less terms - eg. fits to $[S_k/(1+z)]^{3/5}$ for the cosmological model.
e.g. 
\be
\eta(m,z) = A \frac{\tanh [\beta_{1}(m-M_0[S_k/(1+z)]^{3/5}]+1}{2} 
\label{eq:effparam2}
\ee
} There are five free parameters in this parametric form. Since we are dealing with a two dimensional data, having five free parameters is not unusual. Though this parametric form seems to be complicated, it is easy to understand its components. Parameter $A$ sets the overall efficiency amplitude and  $\beta_1$ sets how rapidly the transition between low efficiency and high efficiency should occur. $\beta_2$, $\beta_3$ and $\beta_4$ relating the efficiency to mass and redshift. In particular $\beta_3$ and $\beta_4$  decide if the transition mass should be changed going to the higher redshifts. So by defining this parametric form which is based on our logical expectations, we give enough and reasonable flexibility to the data to be realized in a context of a theoretical model. In Figure~\ref{fig:schem} we see a schematic form of the efficiency function for some arbitrary choices of the free parameters.

\section{Likelihood Analysis and simulations}

Theoretically, if we know the efficiency function $\eta(m,z)$ and the expectation number from the assumed cosmological model $n(m,z,\Delta m,\Delta z)$, we would expect to observe $E(m,z,\Delta m,\Delta z)$ number of clusters in $f_{sky}$ fraction of the sky in $m \pm \frac{\Delta m}{2}$ and  $z \pm \frac{\Delta z}{2}$ intervals:

\be
E(m,z,\Delta m,\Delta z) = n(m,z,\Delta m,\Delta z) \eta(m,z) f_{sky}
\label{eq:effwhole}
\ee

\begin{figure*}[!t]
\centering
\begin{center}
\vspace{-0.05in}
\centerline{\mbox{\hspace{0.in} \hspace{2.1in}  \hspace{2.1in} }}
$\begin{array}{@{\hspace{-0.3in}}c@{\hspace{0.3in}}c@{\hspace{0.3in}}c}
\multicolumn{1}{l}{\mbox{}} &
\multicolumn{1}{l}{\mbox{}} \\ [-2.8cm]
\hspace{-0.3in}
\includegraphics[scale=0.42, angle=0]{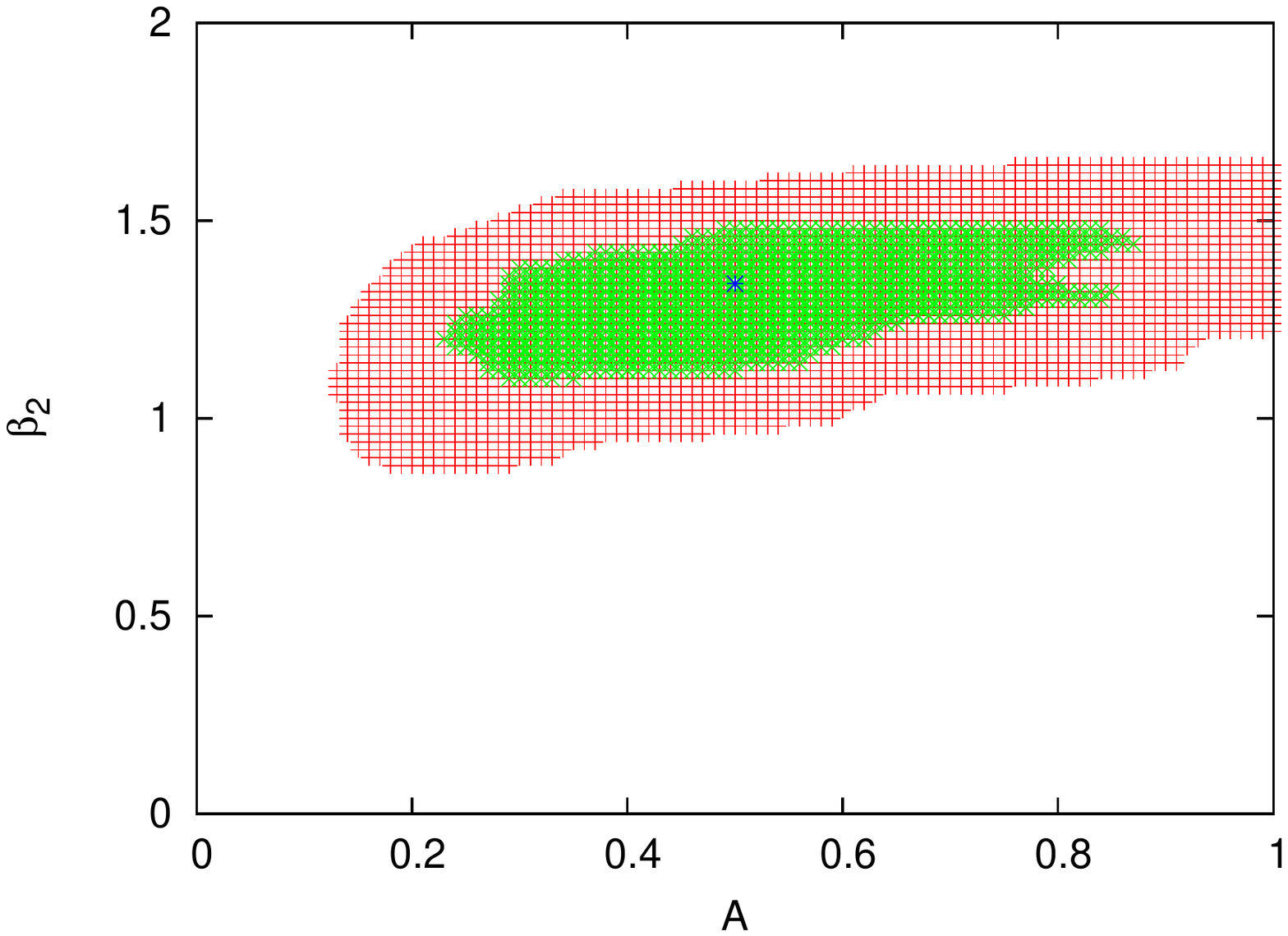}
\hspace{-1.5in}
\includegraphics[scale=0.42, angle=0]{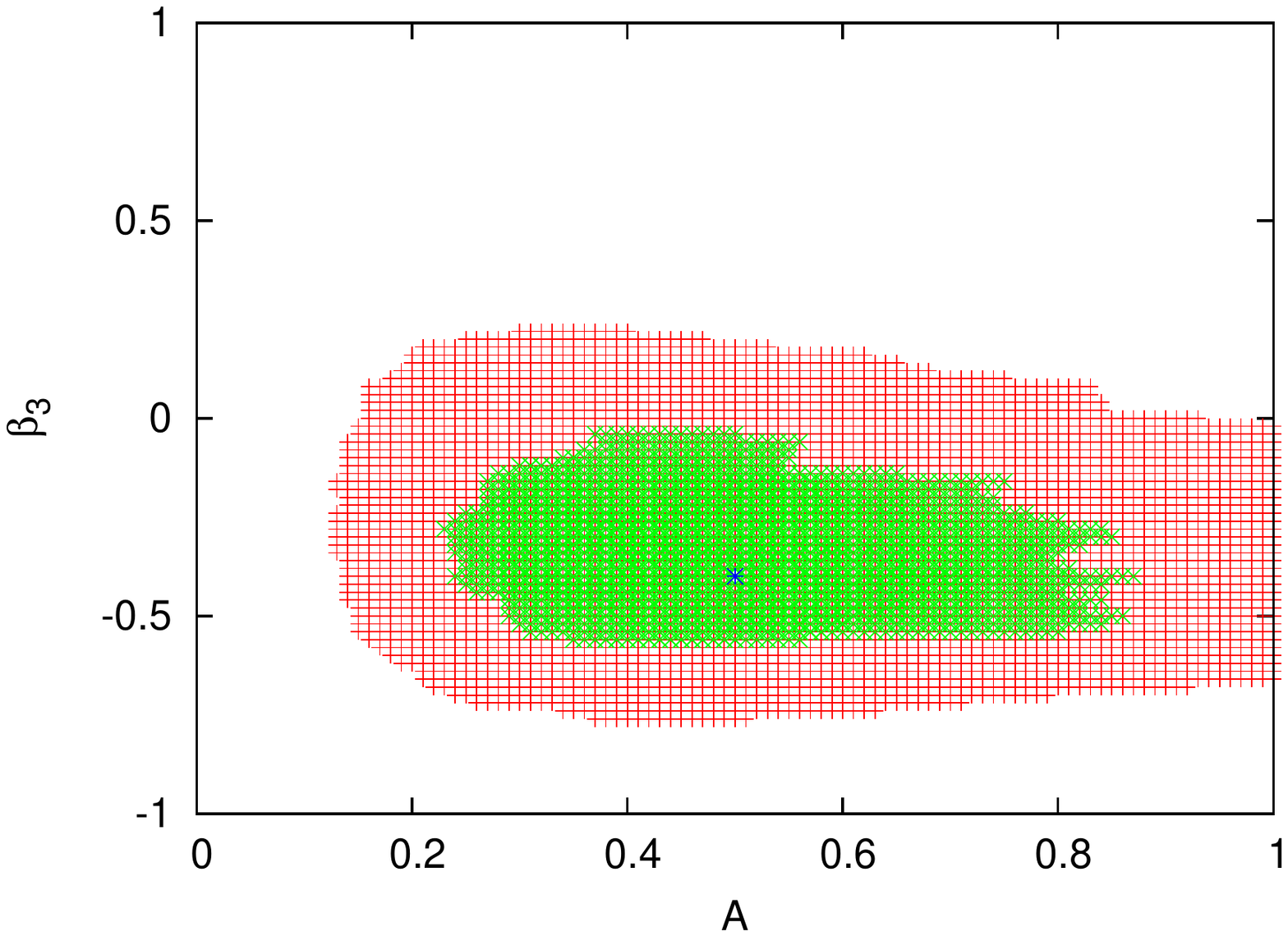}
\hspace{-0.8in}
\end{array}$
\vspace{-0.7in}
\end{center}
\caption{\small  Confidence limits of the three parameters of the efficiency function, $A$, $\beta_2$ and $\beta_3$  around the best fit point (blue dot)  with $\Delta  \hspace{1 mm} Log P < 1$ (green area) and  $\Delta  \hspace{1 mm} Log P < 3.5$ (red area). We put an upper limit for the parameter $A$ as the efficiency function cannot be larger than one.}  
\label{fig:effparam}
\end{figure*}

Since the data has error bars on mass (we neglect the small errorbars on redshifts), we should consider these error-bars in our analysis. These errorbars put uncertainties on the actual position of the clusters in the mass-redshift space. The error-bars on mass of the clusters are usually Gaussian and we can modify Eq.~\ref{eq:effwhole} to the derive $E_{err}$ for each individual cluster:

\ber
E_{err}(m,z,\Delta m,\Delta z) = \hspace{30 mm} \nonumber \\
\frac{\sum_{m'} E(m',z,\Delta m',\Delta z) exp[-\frac{(m-m')^2}{2 \sigma_m^2}]}{U} \nonumber \\
U= \sum_{m'} exp[-\frac{(m-m')^2}{2 \sigma_m^2}] \hspace{20 mm}
\label{eq:Ecomp}
\eer
where $m$, $z$ and $\sigma_m$ are the observed quantities, $E(m',z,\Delta m',\Delta z)$ is given by Eq.~\ref{eq:effwhole}, $U$ is the normalization factor and summation is over the whole mass space.

The probability of observing $r$ events when the mean expected number of events are $E$ is given by Poisson distribution: 

\be
p_{i}(r_i,E_i) = \frac{e^{-E_i}{E_i}^{r_i}}{r_i!}
\label{eq:poiss}
\ee
where in our case, $r_i$ is the number of clusters observed in $i_{th}$ square-interval of mass-redshift and $E_i$ is the expectation number estimated in Eq.~\ref{eq:effwhole} (or Eq.~\ref{eq:Ecomp}). Use of the Poisson distribution is justified as long as our two dimensional bins are sufficiently small.  We now define a joint probability factor or likelihood for the whole data set. For any square-interval of mass-redshift we can calculate the probability of observing zero, one or any number of clusters. Multiplying all these probabilities together we derive the combined probability of the observed data in their particular positions in mass-redshift space:


\be
P= \prod_{all}  p_{i}(r_i,E_i)
\label{eq:prob}
\ee 
since our intervals are small, at most there is only one cluster in each interval. Hence by separating the intervals containing a cluster and those without any cluster we can derive:

\be
P=\left\lbrack\prod e^{-E_i}\right\rbrack_{no\hspace{1 mm}data} \times \left\lbrack\prod e^{-E_i} E_i\right\rbrack_{data}
\label{eq:prob2}   
\ee
simplifying this equation we derive:



\ber
Log P= \left\lbrack\sum Log E_i\right\rbrack_{data} - a N \nonumber \\
N= \left\lbrack\sum E_i\right\rbrack_{all} \hspace{20 mm}
\label{eq:prob3}   
\eer
 where $a=Log(e)= 0.4342944819$ and $N$ acts like a normalization factor. In the likelihood expression given in Eq.~\ref{eq:prob3} probabilities of all intervals even those containing no data are considered. This makes a proper balance between the efficiency parameters and the relative abundance of the clusters. In fact we cannot simply increase the efficiency arbitrarily or generally to large values to get a high likelihood because by doing this we may increase the likelihood for the intervals containing the data (first term in the right hand side of Eq.~\ref{eq:prob3}) but the normalization factor (second term in the right hand side of Eq.~\ref{eq:prob3}) would be also increased. So to find the best fit form we vary the parameters of the efficiency function and we maximize the likelihood.

To do a meaningful statistic we need to estimate the probability distribution function (PDF) of $Log P$ (Eq.~\ref{eq:prob3})) given the $E_i$ distribution and the number of data points. If $E_i$ be close to a Gaussian distribution it is fairly easy to derive the PDF of the $Log P$ analytically as a function of the free parameters of the $E_i$ and number of data points. 
However, our $E_i$ distribution here is a complicated product of the background cosmology expectation $n(m,z,\Delta m,\Delta z)$ and the efficiency function  $\eta(m,z)$ which makes it difficult (if not impossible) to estimate the PDF of the $Log P$ analytically. Thanks to computational advancements it is fairly easy to estimate the form of the probability distribution function using Monte Carlo simulations. Looking at Eq.~\ref{eq:prob3} we also realize that while the $N$ is independent of the number of data points, $\left\lbrack\sum Log E_i\right\rbrack_{data}$ term is directly related to the quantity of the data. So to compare the behavior of $Log P$ for different data sizes we introduce the normalized likelihood function: 

\ber
Log P_{norm}= \frac{1}{n}\left\lbrack\sum Log E_i\right\rbrack_{data} - a N \nonumber \\
N= \left\lbrack\sum E_i\right\rbrack_{all} \hspace{20 mm}
\label{eq:probnorm}   
\eer
where $n$ is the number of data points in our catalogue. We use this normalized likelihood function to compare the results when we have data with different sizes. Now, while we have set our formalism to estimate the likeliness of the data given the back ground theoretical expectation and the efficiency of observing the clusters, and just before applying our method on the real data we test the performance of our approach using simulated data. It is an important task to check if we can trust the proposed method and if it works properly. To do this we fix the efficiency function by choosing some arbitrary values for the efficiency parameters, here we set $\beta_1=100$, $\beta_2=1.1$, $\beta_3=-0.2$ and $\beta_4=0$ and we simulate 1000 realizations of the data. Note that the back ground theoretical model is fixed through out this paper as we have discussed in Section 2. Then we try to use our method to derive back the fiducial parameters of the efficiency function for different data sizes. For simplicity we assumed we know the values of  $\beta_1=100$ and $\beta_4=0$ so we deal with only two actual free parameters that we need to recover, $\beta_2$ and $\beta_3$. Parameter $A$, the overall amplitude of the efficiency function is another free parameter which is sensitive to the size of the data and also to the theoretical expectation which is itself related to $\sigma_8$. So simply, having fixed the theoretical expectation if we have more clusters in our sample, the value of $A$ would be larger. However, $A$ is not sensitive to the relative abundance of the clusters within the data samples and it can be dealt somehow like a nuisance parameter. Results are shown in Figure~\ref{fig:Sim}. Top-left panel shows the relative frequency of the derived $Log P$ from Eq.\ref{eq:prob3} for 1000 realizations of the data with different sizes. In top-right panel we can see the relative frequency of the normalized likelihood for different data sizes using Eq.~\ref{eq:probnorm}. It is clear that by having a larger number of data points, the probability distribution of the normalized likelihood shrinks closer to a mean value. As expected with more numerous samples the log likelihood approaches a Gaussian distribution and the standard deviation of the normalized log likelihood decreases as one over square root of the number of samples. The variance of the log likelihood (not normalized) will increase as the square root of the number of samples. This is to be expected from the central limit theorem as the moments of the $E_i$'s distribution are finite as is the distribution for $Log(E_i)$ and thus the mean will be distributed normally. To use this formalism on a much large sample number, one must reduce the size of the bins so that there is only one cluster per bin so that our simplified Poisson distribution formula can be used. In the lower panels of Figure\ref{fig:effparam} we see the relative frequency of the recovered values of best fit $\beta_2$ and $\beta_3$ for 1000 realizations of the data with different sizes. One can clearly see that the assumed fiducial model is reconstructed quite accurately using our proposed method.

\section{Results}

Now we use our  method to fit actual cluster data. In this analysis we use the recent South Pole Telescope (SPT) SZ-selected sample of the most massive galaxy clusters (considering their statistical errors) that contain 25 clusters in 2500 square-degree of the sky~\cite{SPT}. It is obvious that one cannot expect much from such a low number of data points but we have set our statistics (based on Poisson distribution) in such a way that is reliable in these cases too. Since the data is up to $z \approx 1$ we set $\beta_4=0$ since up to this redshift we would expect a smooth monotonic behavior for the efficiency and having less free parameters would be helpful to estimate reasonable confidence limits where there are not so many data points.


Interestingly after fitting the data we realized that very large values of $\beta_{1}$ are highly preferred. A very large value of $\beta_1$ suggests that a step function can work well in this case and one does not need a sophisticated $\tanh$ function. So setting $\beta_1=\infty$ (practically a very large number to push the $\tanh$ function to one of its limits as a step function) we will end up with a parametric form with only three free parameters (note that we have initially set $\beta_4=0$):

\ber
\eta(m,z)= 0 \hspace{2 mm}; \hspace{2 mm} m-\beta_{2}+\beta_{3}z < 0 \nonumber \\  
\eta(m,z) = A \hspace{2 mm} ; \hspace{2 mm} m-\beta_{2}+\beta_{3}z > 0  
\label{eq:effparammod}
\eer
where $A$, $\beta_2$ and $\beta_3$ are the three fitting parameters we use. 

 In Figure~\ref{fig:effparam} we see the confidence limits of the three parameters of the efficiency function, $A$, $\beta_2$ and $\beta_3$  around the best fit point (blue dot)  with $\Delta  \hspace{1 mm} Log P < 1$ (green area) and  $\Delta \hspace{1 mm} Log P < 3.5$ (red area). The best fit $Log P$ likelihood is derived for $A=0.52$, $\beta_2=1.35$ and $\beta_3=-0.4$. We should note here that we are not doing a $\chi^2$ analysis so we cannot easily assign the $\Delta \hspace{1 mm} Log P$ to $1\sigma$ or $2\sigma$ confidence limits. In fact to make a meaningful conclusion from our analysis we have to involve Monte Carlo simulations to compare our derived results from the real data with the results from many realizations of the data where we know the actual model and the efficiency function.

\begin{figure*}[!t]
\centering
\begin{center}
\vspace{-0.05in}
\centerline{\mbox{\hspace{0.in} \hspace{2.1in}  \hspace{2.1in} }}
$\begin{array}{@{\hspace{-0.3in}}c@{\hspace{0.3in}}c@{\hspace{0.3in}}c}
\multicolumn{1}{l}{\mbox{}} &
\multicolumn{1}{l}{\mbox{}} \\ [-2.8cm]
\hspace{-0.6in}
\includegraphics[scale=0.5, angle=0]{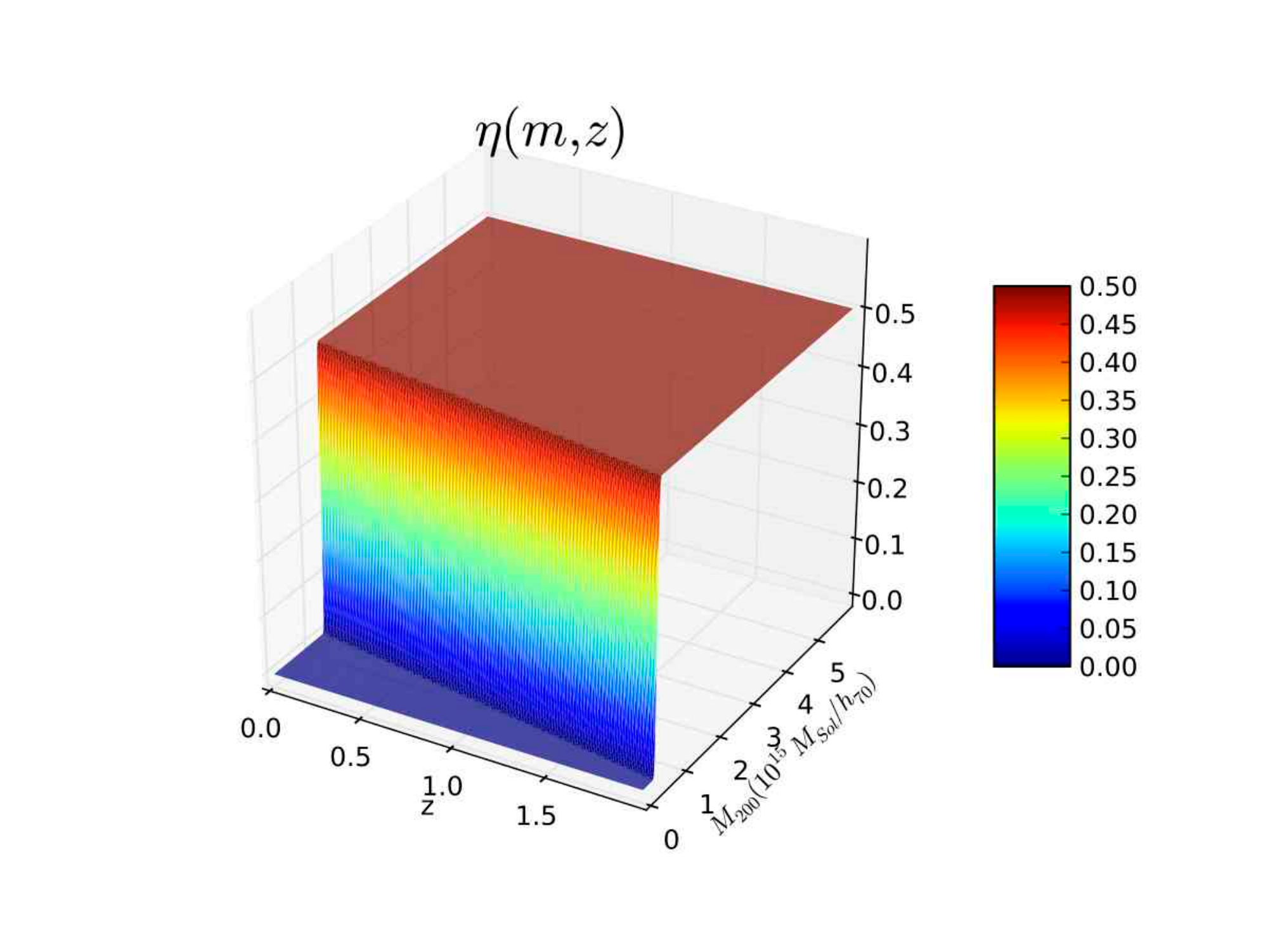}
\hspace{-.6in}
\includegraphics[scale=0.4, angle=0]{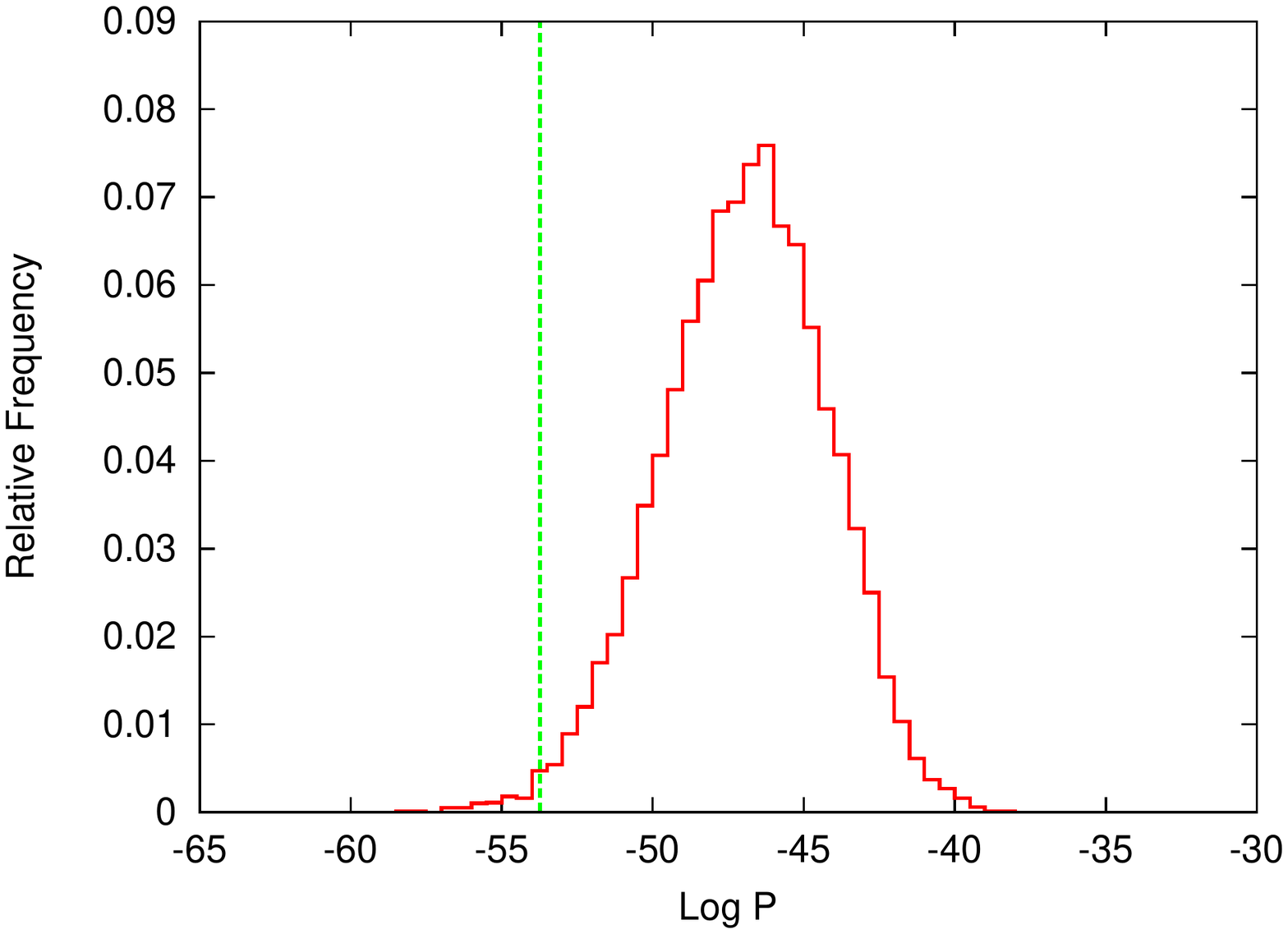}
\hspace{-1.1in}
\vspace{-0.2in}
\end{array}$
\vspace{-0.2in}
$\begin{array}{@{\hspace{-0.3in}}c@{\hspace{0.3in}}c@{\hspace{0.3in}}c}
\multicolumn{1}{l}{\mbox{}} &
\multicolumn{1}{l}{\mbox{}} \\ [-2.8cm]
\hspace{-0.4in}
\includegraphics[scale=0.4, angle=0]{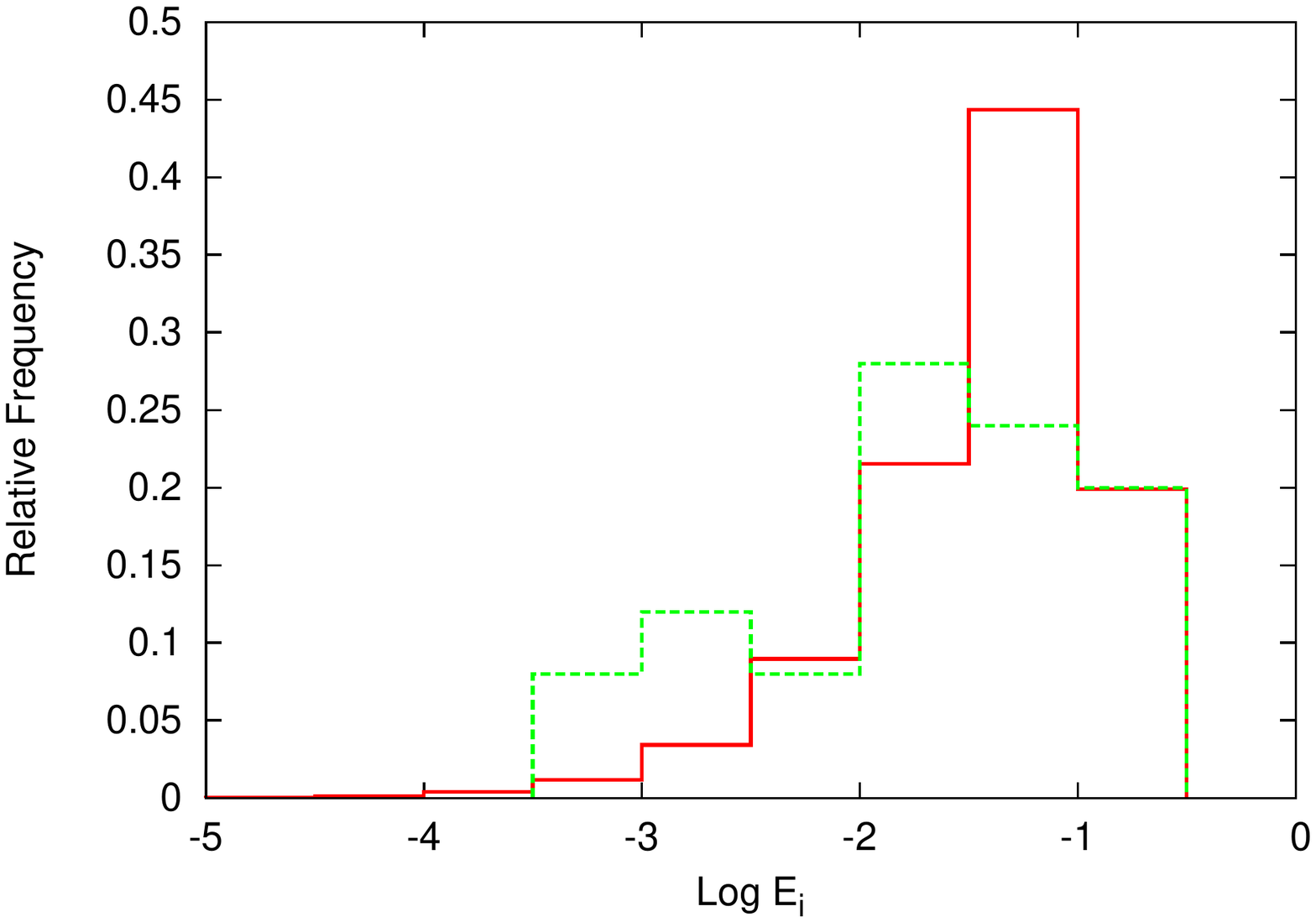}
\hspace{-1.2in}
\includegraphics[scale=0.4, angle=0]{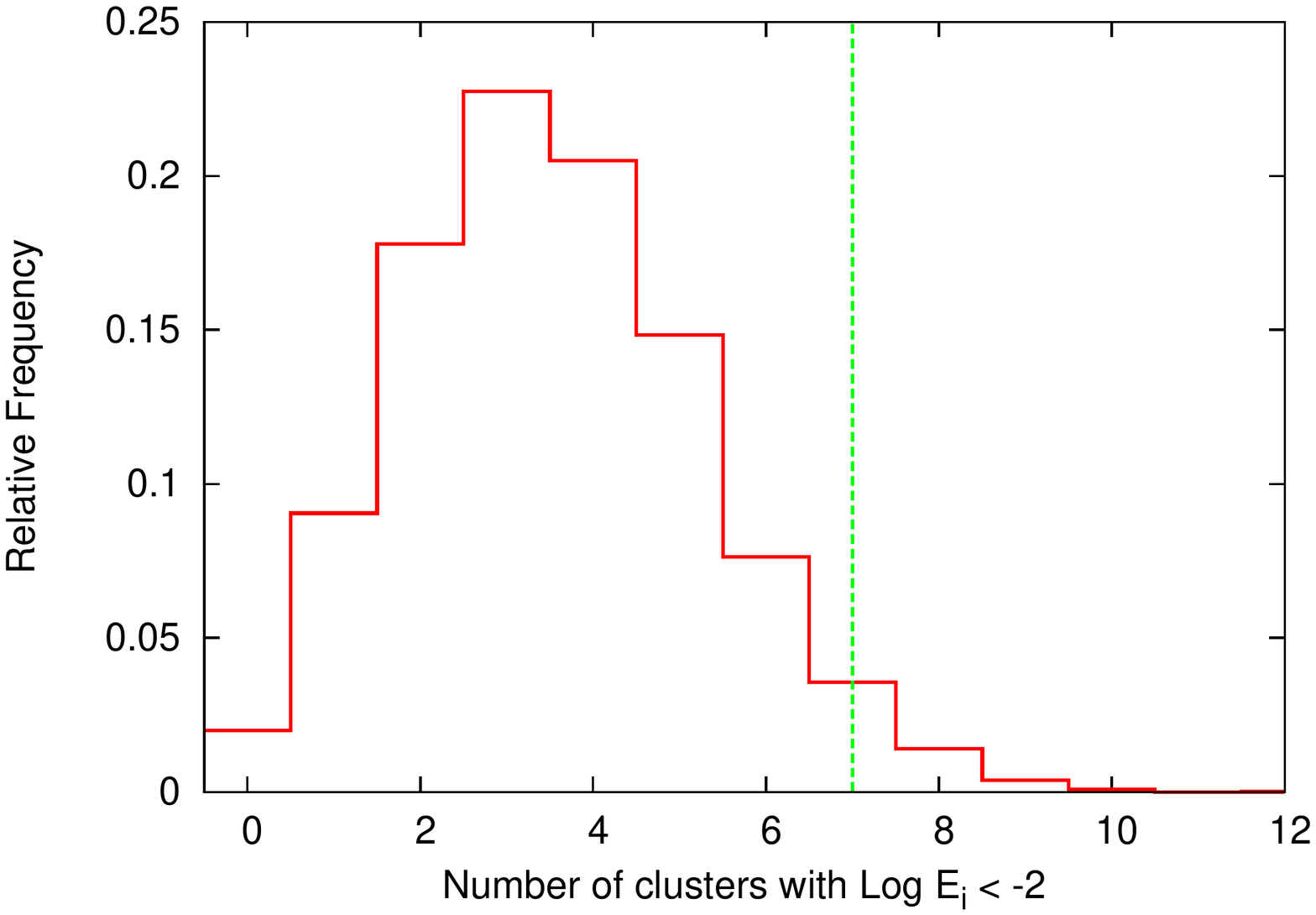}
\hspace{-1.1in}
\end{array}$
\vspace{-0.3in}
\end{center}
\caption {\small Top left panel: Form of the efficiency function for the best fit efficiency parameters with $A=0.52$, $\beta_2=1.35$ and $\beta_3=-0.4$ . Top right panel: Histogram of the derived likelihood for 10000 Monte Carlo realizations of the data in comparison with the derived likelihood from the real data (green vertical line) based on the $\eta$ parameterization given in Eq. 5.1. Lower left panel: Histogram of the $Log E_i$ of the real data points (green line) for the best fit combination of the efficiency parameters in comparison with the histogram of the derived $Log E_i$ for the collection of the individual clusters of the 10000 realizations of the data. Lower right panel: Histogram of the number of clusters with $Log P < -2$ for 10000 realizations of the data in comparison with the case of the actual data (green vertical line). There is a tension between the data and combination of the standard flat $\Lambda$CDM model and the choice of parameterization. Results show about $99\%$ inconsistency between our assumed theoretical model and the data.}  
\label{fig:results}
\end{figure*}

We set $E(m,z,\Delta m,\Delta z)$ according to the assumed cosmological model of flat $\Lambda$CDM model and the efficiency function based on the derived best fit parameters and we generate 10000 realizations of the data. Mass and redshifts of the data points in each set are then derived by two dimensional random Gaussian selection. Now for each simulated data set we do the same procedure as we did for the real data by fitting the parameters of the efficiency function and for the best fit parameters we derive the total likelihood. We can argue that if the derived likelihood from the real data be much smaller than the derived likelihoods from the simulation sets, there must be an inconsistency between the data, theoretical model and the efficiency parameterization. This can be either from a wrong assumption of the background cosmology, in our case the standard Flat $\Lambda$CDM model, or a bad choice of the $\eta$ parameterization which cannot link properly the theoretical model to the observed data. 

\begin{figure*}[!t]
\centering
\begin{center}
\vspace{-0.05in}
\centerline{\mbox{\hspace{0.in} \hspace{2.1in}  \hspace{2.1in} }}
$\begin{array}{@{\hspace{-0.3in}}c@{\hspace{0.3in}}c@{\hspace{0.3in}}c}
\multicolumn{1}{l}{\mbox{}} &
\multicolumn{1}{l}{\mbox{}} \\ [-4.2cm]
\hspace{-0.3in}
\includegraphics[scale=0.42, angle=0]{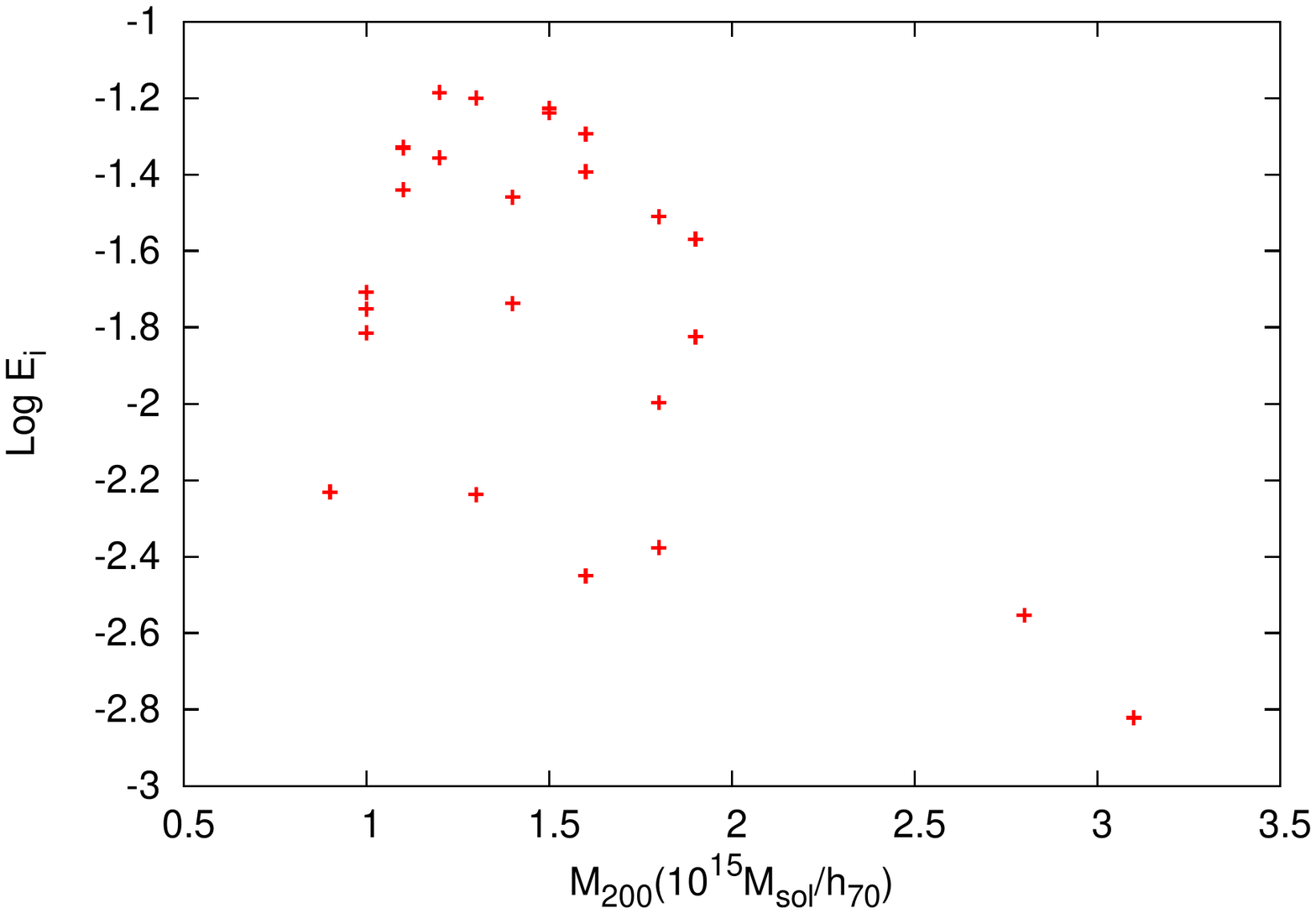}
\hspace{-1.4in}
\includegraphics[scale=0.42, angle=0]{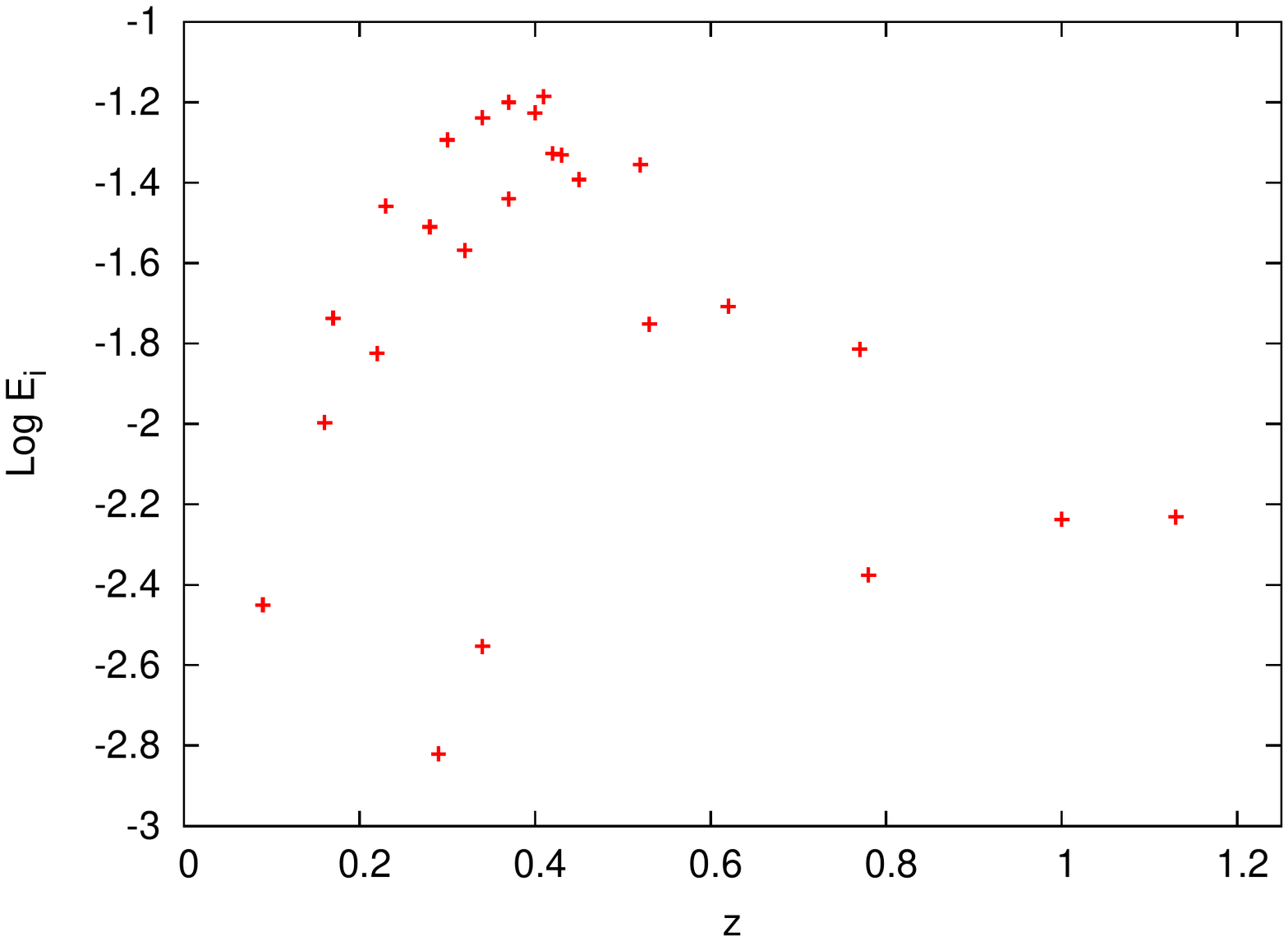}
\hspace{-0.8in}
\end{array}$
\vspace{-0.7in}
\end{center}
\caption{\small $Log E_i$ of the individual cluster data points versus their mass and redshift. One can clearly see that the clusters with the highest mass have a poor likelihood. High redshift clusters also have  lower $Log E_i$  than the lower redshift ones.}  
\label{fig:indiv}
\end{figure*}

Figure~\ref{fig:results} top panel left shows the form of the efficiency function for the derived best fit parameters and right panel shows the relative frequency histogram of the derived likelihood (Eq.~\ref{eq:prob3}) for 10000 realizations of the data in comparison with the derived likelihood from the real data based on the $\eta$ parameterization given in Eq.~\ref{eq:effparammod}. The derived likelihood for the actual data is weakly consistent with the results from simulated data where the assumed model and parameterizations of the efficiency are known and have been assumed correctly. In fact the consistency of the assumed model and the data is only about $P-value = 1\%$ which hints towards a clear tension between the data and our theoretical expectations\footnote{P-value is defined as the probability that, given the null hypothesis, the value of the statistic is larger than the one observed. We remark that in defining this statistic one has to be cautious about a posteriori interpretations of the data. That is, a particular feature observed in the real data may be very unlikely (and lead to a low P-value), but the probability of observing some feature may be quite large – see the discussion in~\cite{haman_shafieloo,shafieloo_crossing}. As a recent example of using a similar approach falsifying cosmological models using other type of data look at~\cite{LP}.}. However we need more data to make any strong conclusion. In fact looking at the likelihood results for individual cluster data points more carefully we see that the clusters with very high mass or those at very high redshifts are indeed responsible for the poor likeliness of the data given the theoretical model and the efficiency parametrization. As we see in Figure~\ref{fig:indiv} the clusters with very high mass (though they are at low redshifts) have the poorest likelihood within our sample. Though error bars on the mass of these clusters are large still because their mean value sits at a very large mass, this makes them quite unlikely to be observed assuming our theoretical model even if the observational efficiency has its maximum possible value. In fact if we just remove the most massive cluster from our sample then the we will have the $P-value \approx 4\%$ which is a reasonable consistency. So we have to be careful in making strong conclusions while the quality and quantity of the data is limited.

While the overall likelihood can be used to test the consistency of the assumed model and the data, it is also important to check if the distribution of the clusters in mass-redshift space allows having a few clusters with a very low $Log E_i$ as we have derived from the actual data. In the bottom-left panel of  Figure~\ref{fig:results} we show the histogram of $Log E_i$ of the real data points for the best fit combination of the efficiency parameters in comparison with the histogram of the derived $Log E_i$ for the collection of the individual clusters of the 10000 realizations of the data. 
In the bottom-right panel we can see the relative frequency histogram of the number of clusters with  $Log E_i < -2$ in 10000 Monte Carlo realizations of the simulated data. 
As we can see it is not extraordinarily unusual to have a random realization of the data with 7 or more clusters having $Log E_i < -2$ which is the case for the actual data.

Our best fit parameters find the observation threshold (efficiency step jump in $\eta$) happens at lower masses for increasing redshift all the way from low redshift.   This is relevant for the mass of the clusters is around $m=10^{15}M_{Sol}$. As we have noted earlier we expect the detection threshold should decrease after $z \approx 1$, but before $z=1$,  we expect an increasing mass threshold with redshift, which is not what the data suggest.   
Of course  $\beta_3=0$ (which represents no change in efficiency as a function of redshift) is less than $\Delta Log P$ of $1$ away from the best fit point but it will be interesting to see in what direction our confidence limit contours will shrink having more cluster data. 
If $\beta_3$ is found to have a considerable negative value, then we have to find a proper explanation for such efficiency behavior or exclude our standard model of cosmology.

\section{Conclusion}

In this paper we study the likeliness of the observed cluster data considering a two dimensional efficiency function for detection of the clusters and assuming the standard Flat $\Lambda$CDM as the background cosmological model. Our analysis is based on the relative abundance of the clusters at different points in the mass-redshift space. Our best fit derived efficiency form of the $\eta$ agrees well with the suggested form estimated by~\cite{SPT}. The overall trend is that the transition in the detection efficiency (from the lower values to the higher values) must be occurred at lower masses in the high redshift in comparison with the lower redshift range (this is related to $\beta_3=-0.4$). Using Monte Carlo simulations we found weak consistency between the standard cosmological model and the observed cluster data using a two dimensional functional form for the detection efficiency. However, due to the small sample of the clusters we used in this analysis, it is difficult to rule out any cosmological model easily by looking at the relative abundance of the clusters. In near future there will be more cluster data available from CMB and SZ surveys which will provide us with a much richer data to test different cosmological models and probably performing the parameter estimation. Detection of few other clusters at high redshifts (they may not need to be necessarily more massive or at higher redshifts than the current ones) or some super massive clusters (even at low redshifts), if we do not observe enough of low mass-low redshift clusters, can change the balance of the observed clusters at different mass-redshifts and make serious problems for the standard model of cosmology. If a model be wrong, it would be hard to assume any simple functional form for the efficiency function $\eta$ to realize the data.

\acknowledgments{A. S would like to thank Reiko Nakajima for the useful discussions. This work has been supported by World Class University grant R32-2009-000-10130-0 through the National Research Foundation, Ministry of Education, Science and Technology of Korea.}


\begin{thebibliography}{99}



\bibitem{SZ_review}
R. A. Sunyaev and Y. B. Zeldovich, Comments on Astrophysics and Space Physics, 2, 66, (1970); R. A. Sunyaev and Y. B. Zeldovich, Comments on Astrophysics and Space Physics, 4, 173 (1972); M. Birkinshaw, Phys. Rep., 310, 97 (1999); J. E. Carlstrom, G. P. Holder and E. D. Reese, ARAA, 40, 643 (2002) 

\bibitem{mortonson}
M. J. Mortonson, W. Hu and D. Huterer, Phys. Rev. D {\bf 83}, 023015 (2011)

\bibitem{SPT}
R. Williamson et al, arXiv:1101.1290


\bibitem{tinker}
J. Tinker et al, Astrophys. J. {\bf 688}, 709 (2008) 

\bibitem{Chamballu2010}
A. Chamballu, J. G. Bartlett and J. Melin, arXiv:1007.3193.


\bibitem{haman_shafieloo}
J. Hamann, A. Shafieloo and T. Souradeep, JCAP, {\bf 04}, 010 (2010). 

\bibitem{shafieloo_crossing}
A. Shafieloo, T. Clifton and P. G. Ferreira, JCAP, {\bf 08}, 017 (2011).

\bibitem{LP}
I. Antoniou and L. Perivolaropoulos, L, JCAP, {\bf 12}, 012 (2010).

 




























\end{thebibliography}
\end{document}